\documentclass[a4paper]{JAC2003}
\usepackage{graphicx}
\usepackage{booktabs}
\usepackage{subfig}
\usepackage{amsmath}
\usepackage{url}

\usepackage{varwidth}
\usepackage{xcolor}

\begin{document}
\title{COHERENT BEAM--BEAM MODE IN THE LHC}

\author{X. Buffat, EPFL, Lausanne, Switzerland; CERN, Geneva, Switzerland\\
R. Calaga, R. Giachino, W. Herr, G. Papotti, T. Pieloni, CERN, Geneva, Switzerland\\
S. White, BNL, Upton, NY, USA}

\maketitle

\begin{abstract}
Observations of single bunch beam--beam coherent modes during dedicated experiments in the LHC are presented. Their role in standard operation for physics is discussed and, in particular, candidates of beam--beam coherent mode driven unstable by the machine impedance are presented.
\end{abstract}

\section{Introduction}
\begin{figure}
 \centering
\includegraphics[width=0.9\linewidth]{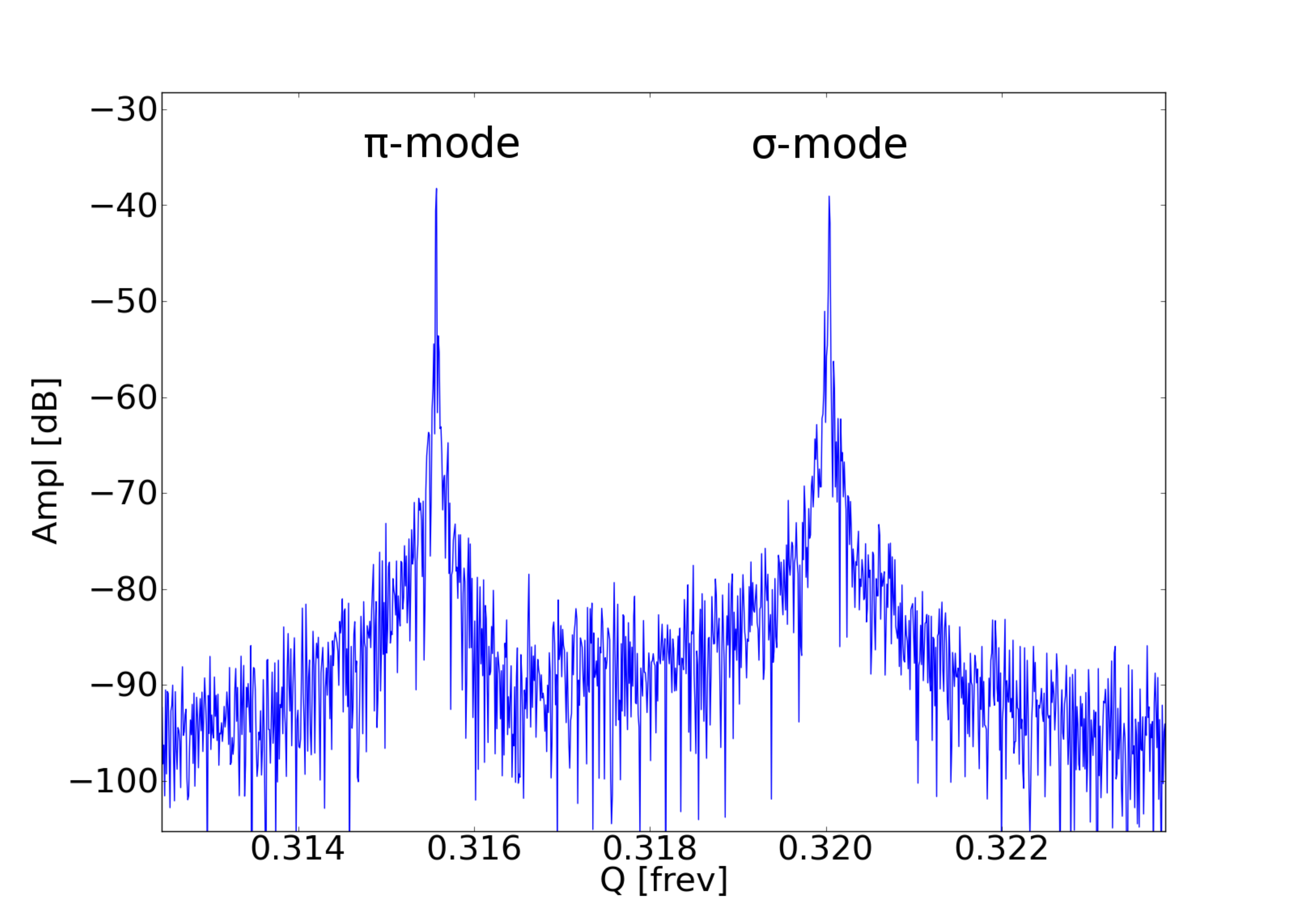}
\caption{Self-consistent simulation of the beam spectrum of two bunches undergoing one HO BB interaction. The $\sigma$-mode lies on the lattice tunes whereas the $\pi$-mode is shifted down by $Y\cdot\xi$, where $Y$ is the Yokoya factor and $\xi$ the BB parameter~\cite{yokoya}.}
\label{fig-COMBI single bunch}
\end{figure}

Two colliding beams are strongly coupled by the Beam--Beam (BB) interaction, be it Head-On (HO) or Long-Range (LR). This coupling can cause the two beams to oscillate coherently in different eigenmodes. When considering one bunch per beam colliding in one Interaction Point (IP), the beams can oscillate in phase, known as the $\sigma$-mode, or out of phase, known as the $\pi$-mode. In such simple configurations, self-consistent tracking simulations with BB and linear lattice transport show that the new eigenmodes, having different frequencies, are not damped in the incoherent spectrum generated by BB (Fig.~\ref{fig-COMBI single bunch}), as already studied in~\cite{stability BB mode}. The complexity increases significantly when considering real LHC cases, with four interaction regions and several LR interactions around each IP.  Previous studies indicate that in such complex configurations, the BB coherent modes tend to be brought inside the incoherent spectrum and are therefore naturally damped~\cite{Tatiana PhD}. These statements are investigated based on observations during dedicated experiments and during luminosity production in the LHC.

Recent studies suggest that coherent BB modes play an important role in the development of impedance driven instabilities~\cite{simon BB2013}. A dedicated experiment aiming at probing this effect is presented.
\section{Stable coherent modes}
\subsection{Dedicated Experiment}
\begin{figure}
 \centering
\includegraphics[width=0.9\linewidth]{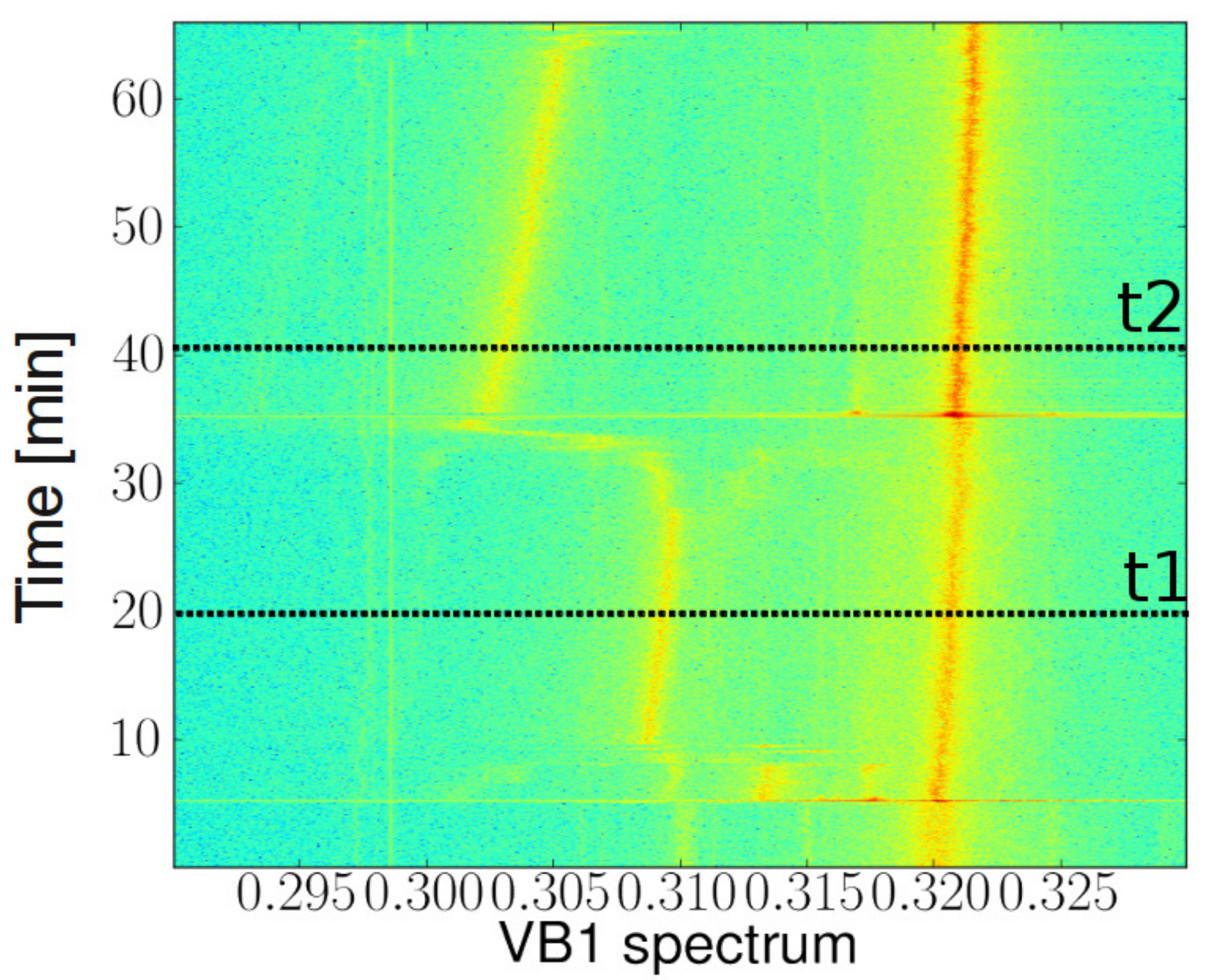}
\caption{Spectrogram in the vertical plane of Beam 1, measured by the BBQ, during an experiment aiming at probing high BB parameters, at injection energy; $\xi\sim0.01$ per IP. At $\sim10$ minutes, the movement of the lower line marks the start of HO collision in IP1, and IP5 at $\sim35$ minutes. The spectrums at $t_1$ and $t_2$ are shown in Fig.~\ref{fig-spectrum HOMD}.}
\label{fig-spectrogram HOMD}
\end{figure}
\begin{figure}
 \centering
\includegraphics[width=1.0\linewidth]{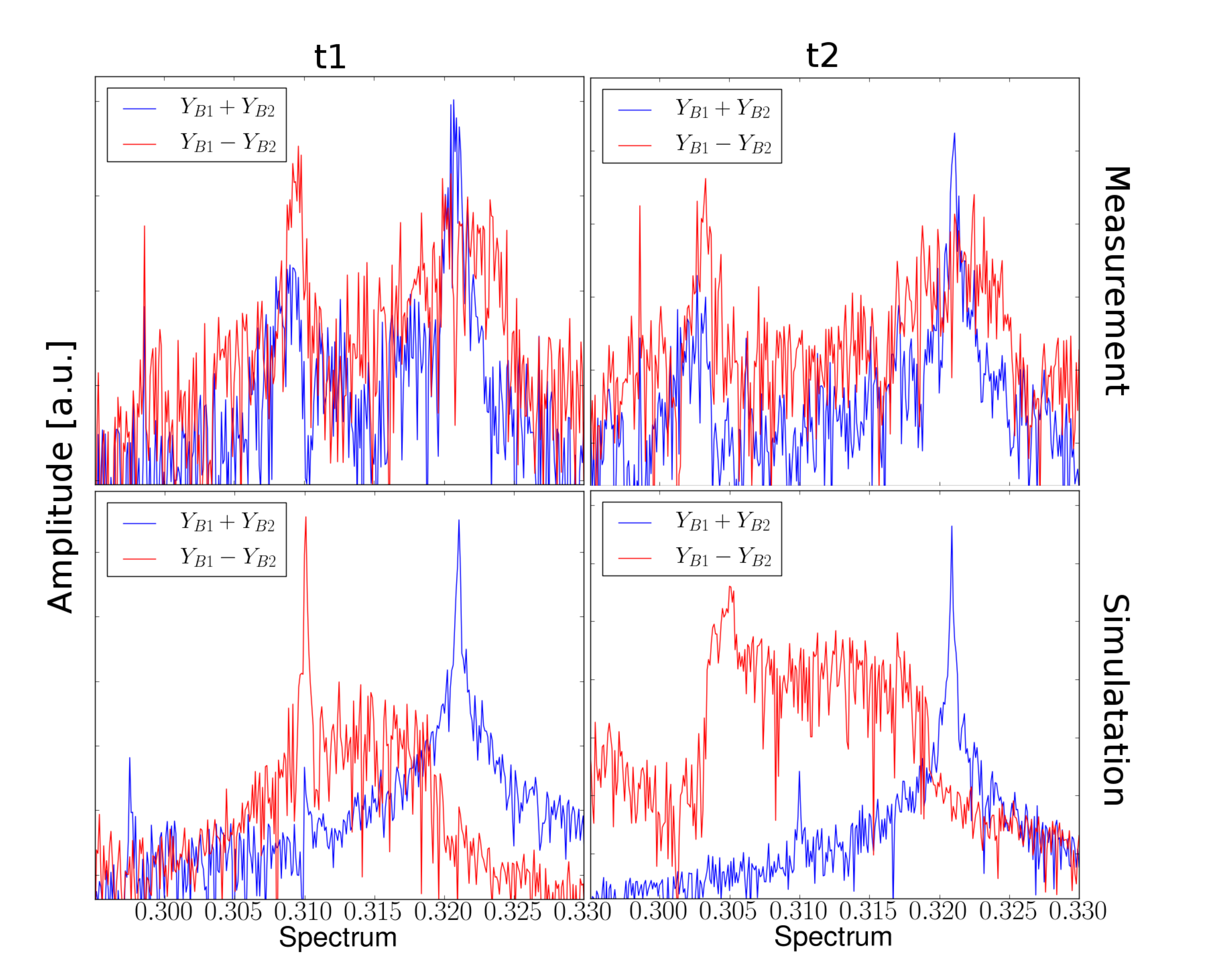}
\caption{The SVD of turn by turn data of both measurement and simulation gives two singular vectors corresponding to the addition (blue) and subtraction (red) of both beams' data, the spectrum of the associated singular vectors are plotted. The two columns correspond to the time defined on Fig.~\ref{fig-spectrogram HOMD}, i.e. one IP colliding and two IPs colliding. The upper row shows measurement from the BBQ and the lower row self-consistent tracking with COMBI~\cite{Tatiana PhD} using measured intensity and emittances.}
\label{fig-spectrum HOMD}
\end{figure}
A series of experiments was performed with single bunches, at injection energy (450 GeV), to probe the maximum BB tune shift achievable in the LHC~\cite{george BB2013}. Some of these tests were done with a single bunch and without transverse feedback. This configuration is ideal to study BB coherent mode, as the Base Band Tune (BBQ) system allows turn by turn measurement of the bunch position. A Singular Value Decomposition (SVD) of this data revealed the presence of BB coherent motion, as indicated by Figs~\ref{fig-spectrogram HOMD} and \ref{fig-spectrum HOMD}. Indeed, not only do the frequencies of the mode observed clearly match the self-consistent simulations, but also the oscillation of the two beams is clearly correlated, in the expected in/out of phase manner. Such behaviour was observed in all experiments performed in this configuration.

Similar experiments were performed, aiming at providing the highest pile-up in the experiments~\cite{george BB2013}, with a few bunches at 4 TeV. Having more than one bunch per beam, turn by turn and bunch by bunch measurement of the position is required to perform an SVD analysis. The pickups used by the transverse feedback can provide such data, as opposed to the BBQ, which cannot. However, the acquisition buffer limits the number of consecutive turns measurable and the sensitivity of these pickups is lower. Nevertheless, they can be used to detect BB coherent mode in dedicated experiments such as these. Unfortunately, a significant coherent signal could not be observed in these experiments, due to the transverse feedback, which, as opposed to previous experiments, was kept on.

\subsection{Luminosity Production}
\begin{figure}
 \centering
\includegraphics[width=0.9\linewidth]{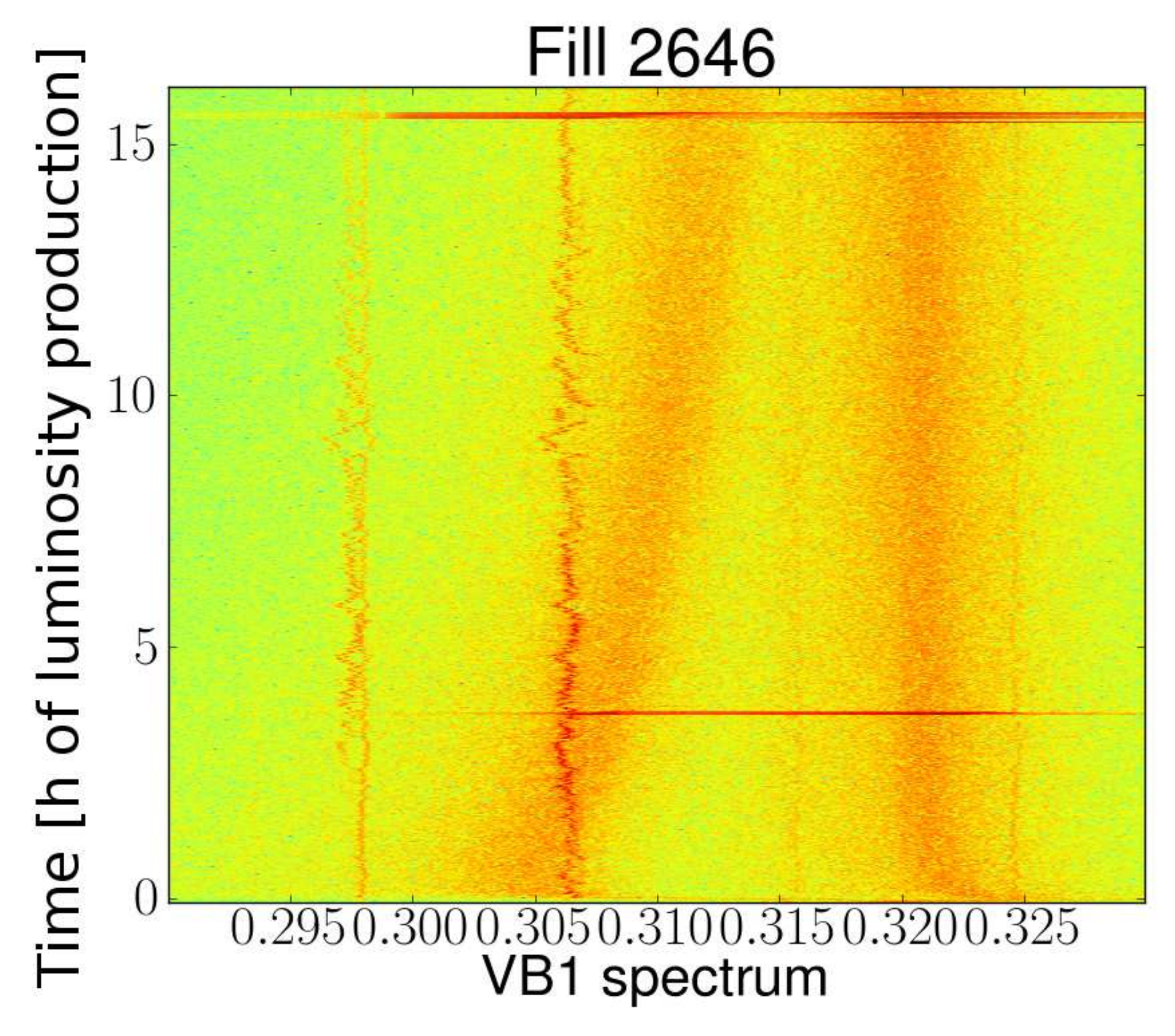}
\caption{Spectrogram in the vertical plane of Beam 1, measured by the BBQ, during luminosity production. The blurred line at $\sim0.321$ is the machine tune $Q_v$, the sharp lines at $\sim0.307$ and $\sim0.298$ are noise lines and the blurred line starting at $\sim0.305$ and moving towards the machine tune is consistent with $Q_v-\xi$, $\xi$ being the BB parameter of the most common bunch, which decays during the fill with the beam brightness.}
\label{fig-spectrogram SB}
\end{figure}
The configuration during luminosity production is very complex; all 1374 bunches are coupled together via either HO or LR interactions in the four interaction regions. Consequently, there exist a variety of modes, with different frequencies, most of them laying inside the incoherent spectrum. Also, the damper is always kept on during luminosity production, which, as mentioned previously, prevents the observation of any coherent mode. For these reasons, no BB coherent modes have been observed during luminosity production. Nevertheless, as shown by Fig.~\ref{fig-spectrogram SB}, there are lines in the BBQ signal at frequencies consistent with both incoherent or coherent motion due to BB interactions. However, without further diagnostics, it is not possible to distinguish between the two.
\section{Unstable coherent modes}
\begin{figure}
 \centering
\includegraphics[width=0.9\linewidth]{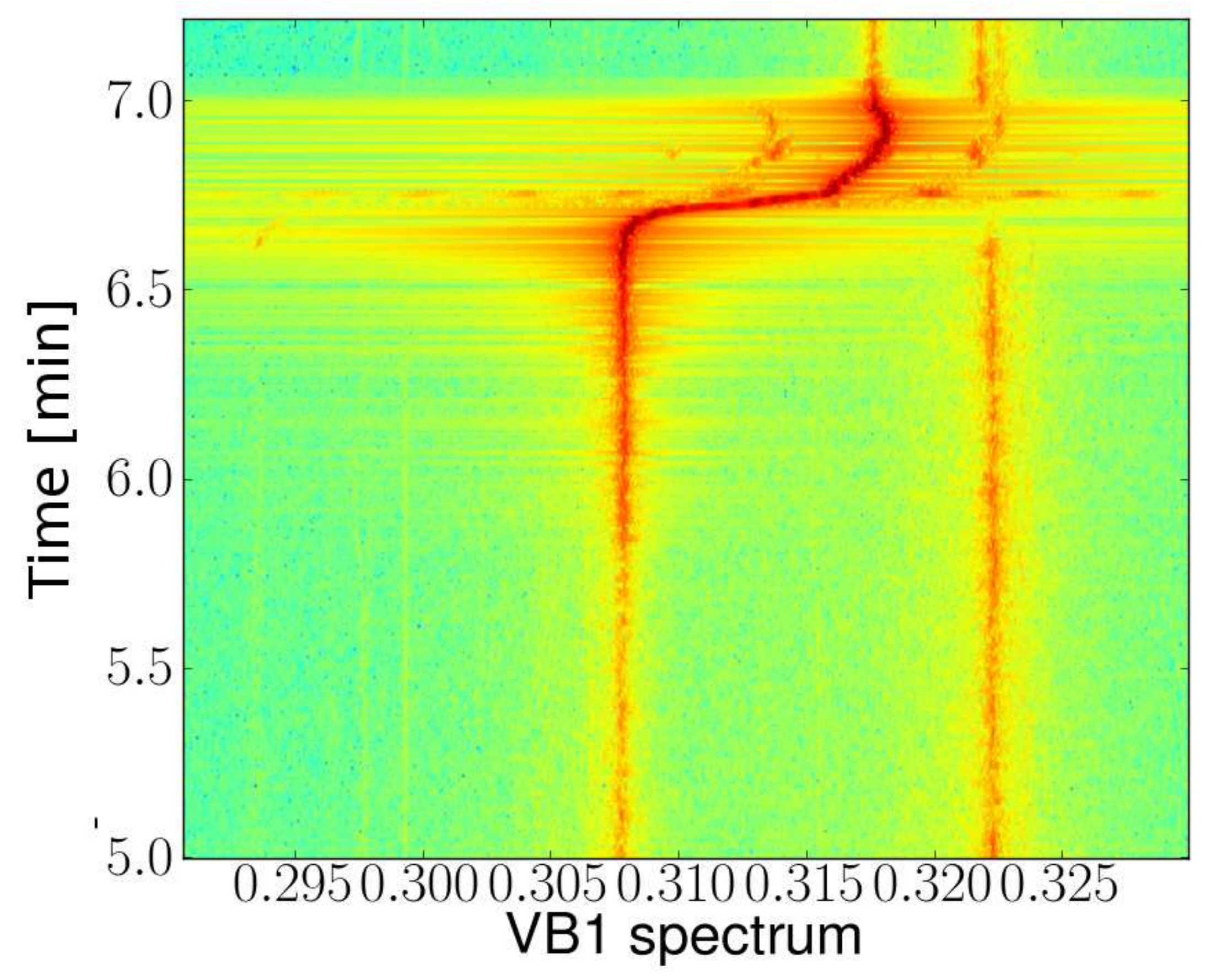}
\caption{Unstable $\pi$-mode observed during a dedicated experiment aiming at probing large HO BB parameters at injection energy with one high brightness bunch per beam colliding in IP1\&5.}
\label{fig-HOMD instability}
\end{figure}

At the end of the experiment presented in Fig.~\ref{fig-spectrogram HOMD}, a coherent mode, previously demonstrated to be a $\pi$-mode, became unstable (Fig.~\ref{fig-HOMD instability}). The beams stabilize themselves naturally at the expense of large intensity loss and emittance growth; the frequency of the mode after the instability reflects the reduction of beam brightness. Beam--beam modes are not self-excited; in this case, the driving force is unknown. In particular, the lack of chromaticity measurement during the experiment and the large uncertainties of the LHC impedance model at injection energy~\cite{impedance model} do not allow quantitative comparison with models including BB and impedance. Such instability was never observed with colliding beams at top energy. 

During luminosity production, there have been plenty of observations of instabilities when the beams were colliding with a transverse offset, despite the presence of a strong transverse feedback. In some cases, the instability was observed in one beam only and could be explained by a lack of Landau damping of pure impedance mode~\cite{single beam instability}. In some others, the instability is observed on both beams. These observations could be consistent with a single beam instability going with an incoherent transmission of the signal to the other beam. Nevertheless, a coherent beam--beam instability is not excluded.

\begin{figure}
 \centering
\subfloat[]{
\includegraphics[width=0.9\linewidth]{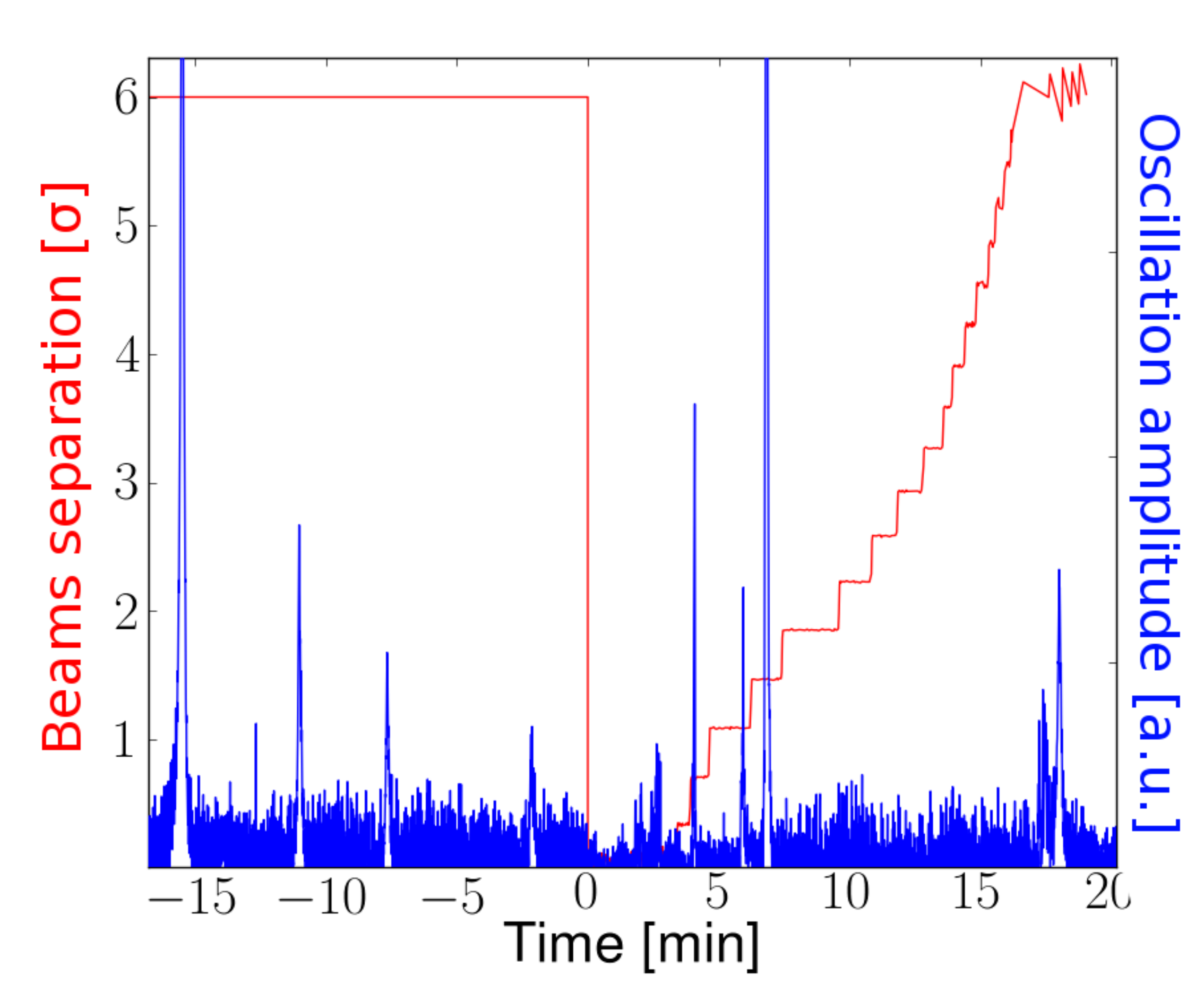}
\label{fig-eofMDMD BBQampl}
}
\qquad
\subfloat[]{
\includegraphics[width=0.85\linewidth]{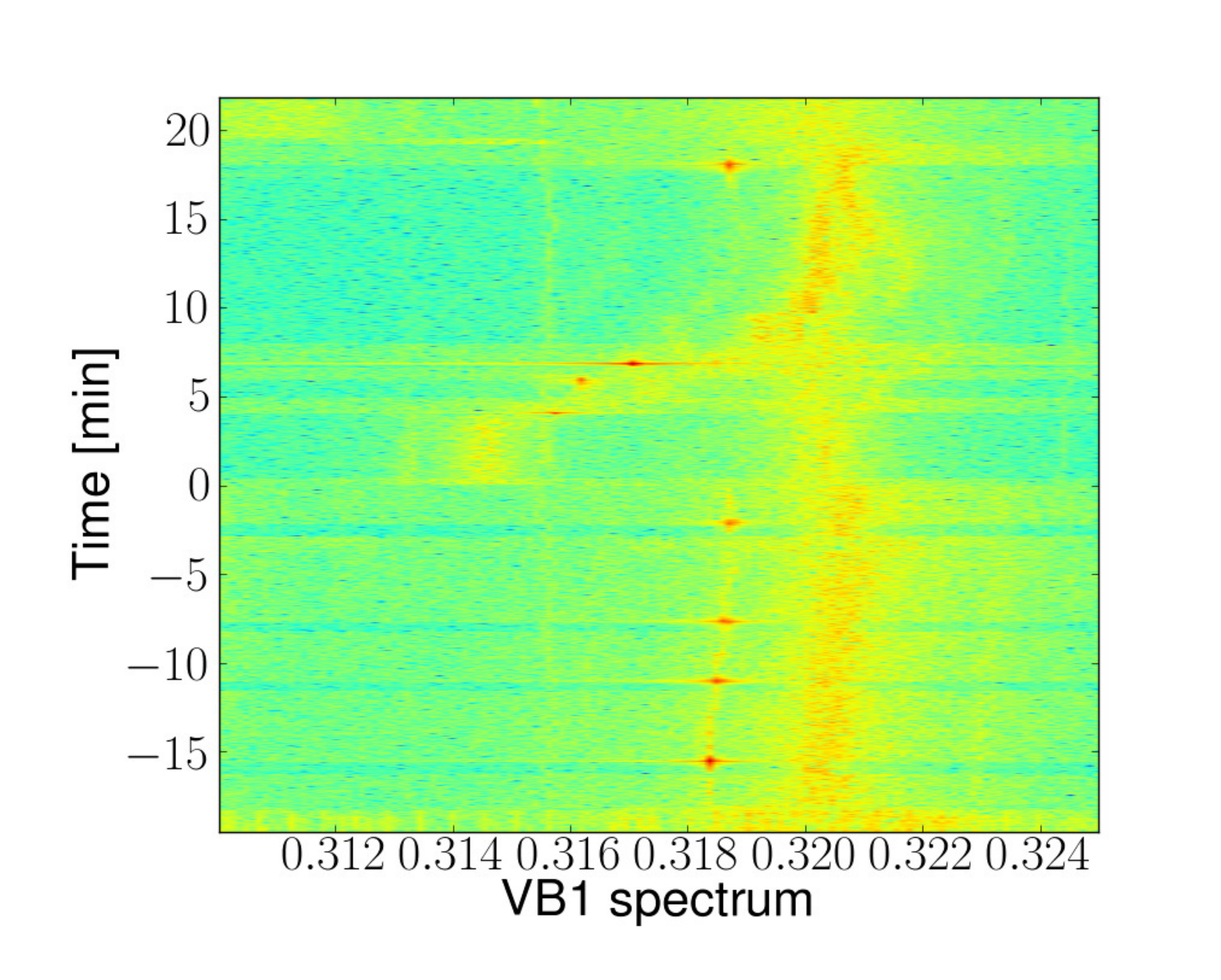}
\label{fig-eofMDMD spectrogram}
}
\caption{Spectrogram, vertical plane of Beam 1.}
\label{fig-eofMDMD}
\end{figure}

\begin{figure}
 \centering
\subfloat[The measured rise time is 5.9~s at 6~$\sigma$ separation. Only Beam 1 is unstable in the vertical plane.]{
\includegraphics[width=0.9\linewidth]{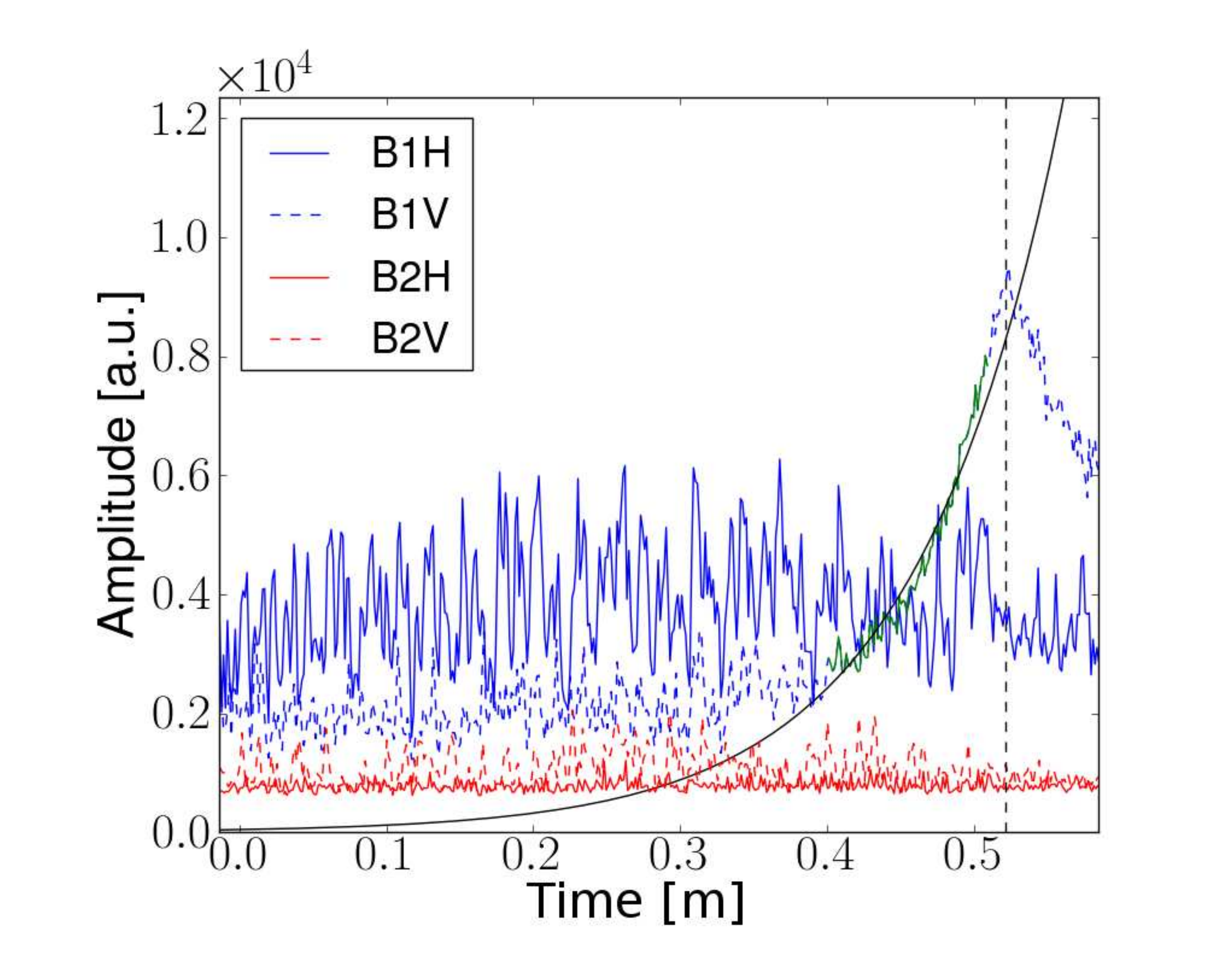}
\label{fig-eofMDMD risetime 1}
}
\qquad
\subfloat[The measured rise time is 1.8~s at 1.4~$\sigma$ separation. Both beams are unstable in the vertical plane.] {
\includegraphics[width=0.9\linewidth]{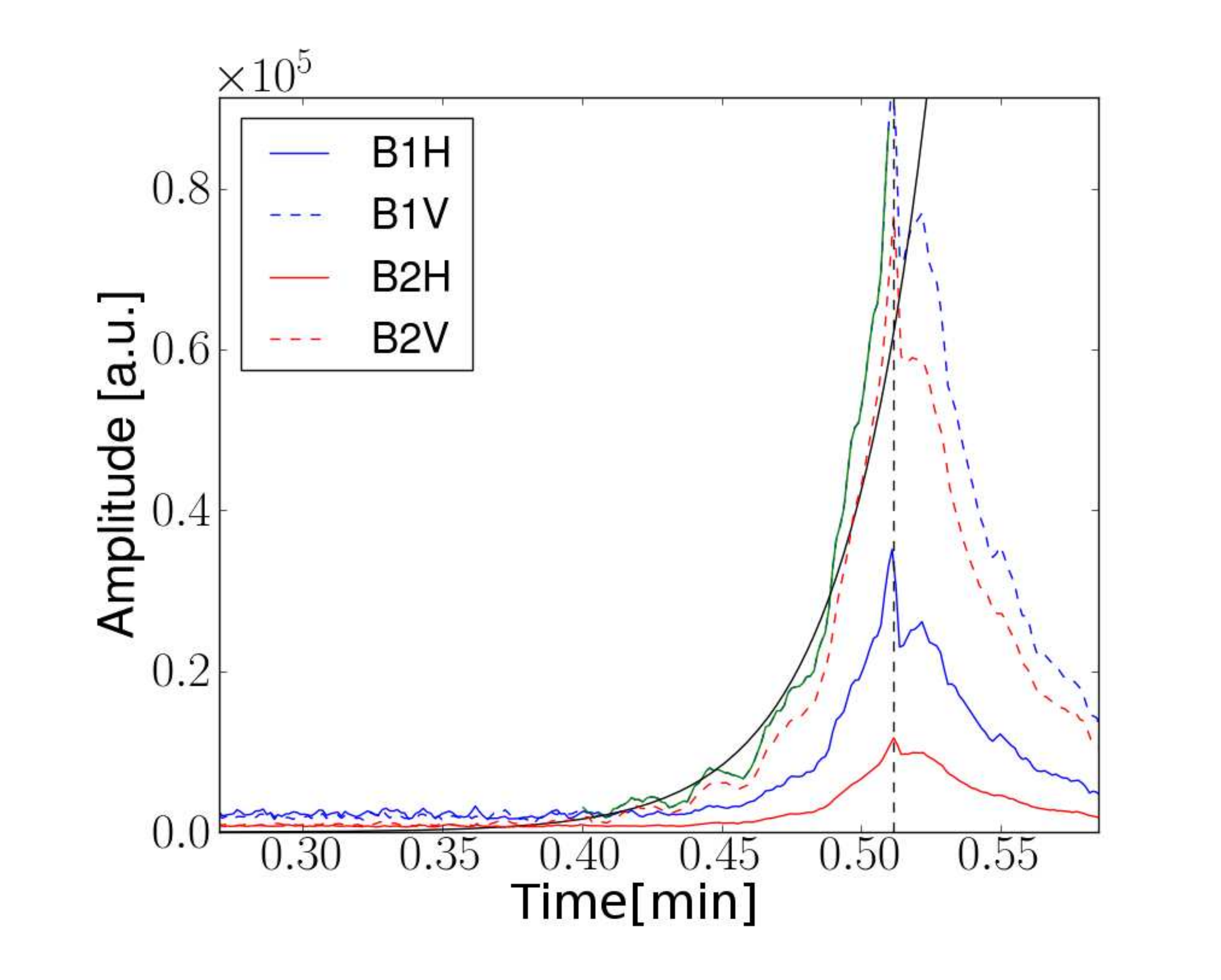}
\label{fig-eofMDMD risetime 2}
}
\caption{Measured oscillation amplitude in both planes of both beams, with an exponential fit to measure the rise time. The instability is damped before a significant degradation of the beam brightness by turning the transverse feedback on, marked by a vertical dashed line.}
\label{fig-eofMDMD risetime}
\end{figure}

These observations motivated theoretical developments and a dedicated experiment at the end of a special fill~\cite{eofMDMD}. After the squeeze, two bunches per beam were colliding in IP1\&5. First, the beams were re-separated by 6 $\sigma$, the r.m.s. beam size, in both IPs and the chromaticity was set to $\sim5$ units. Before $t=0$, in Fig.~\ref{fig-eofMDMD}\subref{fig-eofMDMD BBQampl}, a series of spikes in the oscillation amplitude mark a few tests of the beams' stability without transverse feedback, by turning it off and on again when an instability is observed. The octupole strength is increased after each tentative; it was found that the maximum strength was not sufficient to stabilize the beams. At $t=0$, the beams were brought into collision in IP5 with the feedback on. Once the beams were colliding HO, the feedback was no longer required to maintain the beams' stability. The beams were then re-separated in steps, visible in Fig.~\ref{fig-eofMDMD}\subref{fig-eofMDMD BBQampl}. At each step, the stability without transverse feedback is tested, as previously. It was observed that the beams are stable for separations below 0.7~$\sigma$ and from 1.8 to 6~$\sigma$, whereas unstable from 0.7 to 1.8~$\sigma$ and at 6~$\sigma$. Also, the instabilities at intermediate separations have different characteristics than for the 6~$\sigma$ separation. As shown by Fig.~\ref{fig-eofMDMD risetime}, at intermediate separations both beams are unstable, as opposed to 6~$\sigma$. Moreover the rise times are significantly different (Table~\ref{tab-risetime}).

\begin{table}
\caption{Measured rise time at different separations}
\centering
 \begin{tabular}{cc}
\hline\hline
Full separation [$\sigma$] & Rise time [1/s] \\
\hline
0.7 & 2.7 \\
1.1 & 6.7 \\
1.4 & 1.8 \\
6 & 5.9 \\
\hline\hline
 \end{tabular}
\label{tab-risetime}
\end{table}

Due to the lack of time, the scan in separation is extremely coarse; moreover, important parameters, such as chromaticity and emittances, are poorly know. These factors render a quantitative comparison with theoretical models difficult. Nevertheless, the existence of a critical separation, in the order of 1 to 2~$\sigma$, is in accordance with a lack of Landau damping due to the modification of the tune spread caused by the beam--beam force~\cite{single beam instability}, as well as a coupled beam--beam and impedance mode~\cite{simon BB2013}. The experimental evidence does not allow us to distinguish between these two models, which are not mutually exclusive. Nevertheless, the fact that both beams became unstable simultaneously, with identical rise times (Fig.~\ref{fig-eofMDMD risetime}\subref{fig-eofMDMD risetime 2}) is a indication that the two beams oscillate coherently.

It is important to stress that, as predicted in~\cite{simon BB2013}, the transverse feedback was efficient at stabilizing the beams. Indeed, during this experiment, no instabilities were observed while the transverse feedback was active. This is different with respect to what is observed during luminosity production, as instabilities are observed while the transverse feedback is on. The configuration is nevertheless very different, in particular the presence of multiple bunches is expected to have a strong impact on the dynamic. The models are therefore being extended to assess the full LHC complexity in the multibunch regime.
\section{Conclusion}
Beam--beam coherent modes have been observed in the LHC, during experiments with single bunches and without transverse feedback. Their frequency and the corresponding eigenvectors follow theoretical models and simulations. Some frequencies observed in the beam spectrum, while running with multibunch and with the transverse feedback on, could be attributed to coherent beam--beam modes. Nevertheless, the instrumentation available could not measure the correlation between the beams, and thus do not allow demonstration of the presence of coherent motion.

Recent developments suggest that the stability of beams colliding with a transverse offset can be critical. An experiment was performed, the results being in qualitative agreement with the models. In particular, it was shown that beams, being unstable when separated, can be stabilized by HO collision, removing the need for the transverse feedback in this configuration. Also, it was shown that there exists a critical separation at which the beams' stability is reduced. The two beams were strongly coupled during instabilities observed while colliding with a transverse offset in the order of 1 to 2~$\sigma$. Unfortunately, as in the previous case, the lack of diagnostics  prevents demonstration of the presence of coherent motion.

While the frequency and eigenvectors of the beam--beam coherent modes are well understood and agree well with observations, the stability of these modes, in particular in the presence of the machine impedance, still requires both theoretical and experimental investigations to fully assess the LHC complex configurations.

\end{document}